 \documentclass[11pt]{article}

 \pdfoutput=1

\usepackage{graphicx} 
\usepackage{amsmath}

\usepackage{amssymb}

\usepackage{enumitem}

\usepackage[francais,english]{babel}

\usepackage{float}
\usepackage{ifthen}
\usepackage{hyperref}

\begin{document}

\title{\bf The last paper ``On the theory of storm'' \\ 
(Zur Sturmtheorie) published by \\
Max Margules in 1906.}

\author{A translation by Pascal Marquet. {\it M\'et\'eo-France}. This translation may be improved \\
 and  any comments and/or suggestion of improvements: pascal.marquet@meteo.fr}


\date{\today}

\maketitle



\vspace*{-8 mm}

\noindent\hfil\rule{0.99\textwidth}{.4pt}\hfil
\vspace*{2 mm}

\noindent Beginning of the paper: {M. Margules} {(1906)}.
Zur Sturmtheorie.
{\it Meteorologische Zeitschrift\/} \\
\hspace*{4cm} {\bf Vol. 23} (Issue 11): p.481--497. \\
\hspace*{22mm} \url{https://archive.org/details/meteorologische01unkngoog}
\vspace*{2 mm}


Recently, I have published a study on the energy of the storms\footnote{Note of the translator: the famous ``On the energy of the storms'' by Margules (1903-1905)}, which still needs some additions.
I think I am able to apply in this communication a section that has been elaborated since then.
Circumstances, which I cannot go into here, cause me to write down the lecture and the supplement, \underline{\it and then to bid farewell to meteorology\/}. \footnote{underlined by the translator}.

\section{\underline{Three views on the energy source of the storms}.}
\label{section_I}

\noindent \hspace{3 mm} $\bullet$ {\bf 1.} Long ago it was assumed that, like the regular winds, turbulent air movements 
are caused by horizontal temperature differences and are produced by the 
effect of the gravitational force.
These assumptions are obvious and the reason why they have been abandoned is not quite clear.
I would like to mention some of the experiences that have been gathered in those districts, as other storm theories have been preferred.
In the great storm regions of Europe and North America, two parts of strikingly different temperatures are found close together.
In severe storms of small extent (gusts), sudden temperature falls of 10 degrees and above can be observed.
Tornadoes develop near the border between cold and warm air masses.

If the trade winds are fitted with the example of the constantly heated and the constantly cooled chambers between which the door is open, the same is true of the storms.
The door opens between a warm and a cold room and the air chambers must be thought up a few kilometres high and many hundreds, even a few thousand kilometres long and wide.
It is possible to calculate the amount of {\it kinetic energy available\/} in such a system, or how fast the velocity is, when the vigorous force is distributed uniformly over the whole mass. 
The course and the results of the calculation are reported in the second section.

\vspace{3mm}
$\bullet$ {\bf 2.} Many storms are accompanied by abundant rainfall. 
In numerous writings from the last four decades it is thought that the potential energy of the air mass before the storm is to be found in the vapour content, and that part of the condensation heat is converted into kinetic energy.
From some of the books that I have at hand, I take the following passages:

- Espy (cited in Reye, Hurricane Storms, p.147): All storms are generated by steam.

- Reye (p. 134): The moving force in the hurricanes is that of the heat which is released by condensation of atmospheric water vapour.

- Ferrel (Popular Treatise on the Winds, p. 283): The energy of the cyclone is mostly in the aqueous vapor Condensed.

- Eliot expressed the same opinion in many places of his work on East Indian cyclones.

In the manner in which the heat of condensation is converted into kinetic energy, we obtain the following idea (which is given in Ferrel's textbooks, p.232, 231): a moist air mass rises, it has a higher temperature at each level than the air around, it drifts upward, other moist air recedes, this also keeps the air of a large area in motion.
The source of the energy here is the latent heat of steam, which reduces the cooling of the ascending mass, the motor the gravitational force.

The velocity that an ascending moist air particle in a quiescent mass of given temperature stratification can reach is easy to calculate.
Our task requires the calculation of the {\it available kinetic energy\/} \footnote{made in italic by the translator} of a system in which there is an extensive mass of steam, which is next to (or under) another. 
We have quite long calculations to make, but we cannot avoid the part of the heat of condensation which is converted into {\it living Force\/} (namely twice the kinetic energy, or $V^2 \: dm$? / the translator). 
That a very small part might suffice to produce violent storms has already been noticed by Reye (p.160).

Following the calculation of the two-chamber system, I have investigated a model which contains steam in a chamber instead of the warm dry air. 
This rises, spreads over the sinking cold mass, it condenses by expansion.
But the heat of condensation does not contribute to kinetic energy in such an arrangement.
The {\it available living force\/} is no greater than it would be with dry air of equal density.

There are still systems to be investigated in the manner of those of Ferrel and Helmholtz, where vaporous air passes through dry masses when rising: kinetic energy can be produced at the expense of condensation heat.
The analysis of simple processes of this kind is the only new thing I have to say here. (Section III)

The views outlined in Figures 1 and 2 do not exclude one another.
It would be possible that in the middle and high latitudes, the first, and in the tropics, the second theory, represent the essential conditions of the formation of the storm.

So far we have not mentioned the horizontal pressure gradients.
The meteorologists pay the most attention to this and there are not a few to whom the sentence is sufficient: the steep gradient produces the storm.
But crowded isobars and storms appear simultaneously.
In the storm area, the kinetic energy is far greater, and at the same time a form of potential energy (we may call it tentatively the potential energy of the horizontal pressure distribution), larger than it, was present before.
Both must be at the expense of other forms: the potential energy of the situation and the internal energy (in the latter case, the internal heat of steam is included as well as the potential energy of the horizontal pressure distribution, but it can grow while the remaining part decreases)

\vspace{3mm}
$\bullet$ {\bf 3.} In Hanns Lehrbuch (first edition, p.578) one finds the following about East Indian storms:

``{\it The sources of the energy of the latter are given by the Indian meteorologists in the formation of large hurricanes.
This consists in the fact that the preexisting meteorological conditions make it possible for air masses to flow from all sides, and, to a great extent, to turbulence, to a certain earth point.
A very slight gradient can supply the observed agitation energy under these conditions to the internal ``vertebral rings''.
Since, in the tropics, the distracting force of the earth's rotation is still quite small, this energy can concentrate on a relatively small space and produce the immense wind strengths there\/}.''

This was extensively developed in the Journal of Meteorology of 1877, has passed into the book of Hildebrandsson and Teisserenc de Bort, and others, and appears to be the dominant view now of the energy source of tropical cyclones.

There is no mention of ascending and descending air masses, but as of a favourable circumstance in the following passage of the textbook.

I often had the opportunity to learn from Prof. Hann, and it is always a pleasure for me to agree with him. 
From this comes the duty of gratitude to explain why I consider vertical movements under the influence of gravity to be the essential condition of the storm.

If I do not take vertical flow components, I can imagine the whole process in a thin air disk. 
Given is a certain horizontal pressure distribution in the initial state.
Their working value shall be calculated.

$p'$ shall be the mean pressure, $p' \: (1 + \epsilon)$ the observed before the storm; $\epsilon$ is a small fraction, positive in one part of the field, in the other negative, its surface integral zero; [$\: \epsilon \:$] the mean of the absolute values of $\epsilon$.
The greatest work which may be required of the given distribution of pressure is made up to the equilibrium of the pressure. It is so great that it can produce the velocity [$\: \epsilon \:$]$\: \times \: 236$~m/s in the whole air-disk.
Let us take the value [$\: \epsilon \:$]$ \: = \: 1/760$, which Hann assumes in an example: the velocity of 0.3~m/s is then obtained for the whole mass.
Suppose that in a part of the region the {\it vital force\/} is concentrated, producing the velocity of $30$~m/s, this could only be in the tenth thousandth part of the whole active mass.
For a tropical storm of $100$~km in diameter, the potential energy of an air disk of the diameter of an earth-quadrant would have to be considered.

Note that the potential energy of a horizontal pressure distribution is very small, even with steep gradients.
If one is looking for the energy source of the storms, one hardly needs to worry about the gradients.
In the discussion which has developed from Hann's theory, Sprung has used the very apt designation of ``nourished and unskilled gradients''.
In the last decades these words have fallen into oblivion.
In the air disk, we have unsuccessful gradients, as well as in the atmosphere, if the relative pressure distribution is the same in all levels. 
Such gradients disappear by very small movements which occur in the sense of the gradient.
From the food of the gradients, the condensation heat of ascending steam-rich air currents was almost exclusively assumed.
Recent observations have shown that even in extensive air masses, which have large temperature differences at the same level, sufficient food can be supplied to the gradient without condensation, in order to generate and maintain storms. 

The pressure gradient is only associated with the importance of a distributor of the kinetic energy.
This will be discussed in more detail if better methods than the previously known ones have been found for the treatment of aerodynamic tasks.
Undoubtedly, the greatest velocities arise in the way that air flows in the sense of the horizontal pressure gradient, as Hann has calculated in the cited treatise.

But the source of energy is not to be found in the gradients which grow at the same time as the wind, but in the processes which result from the action of the gravitational force, which feeds the pressure gradient: in descending coldly, ascending, warmer or steamy air.

\section{\underline{The energy equilibrium of dry air}.\protect\footnote{Presentation of some sections of the treatise on the energy of storms. Yearbooks of the Central Institution for Meteor. 1903. (Vienna, 1905.)}}
\label{section_II}

\noindent \hspace{3 mm} $\bullet$ {\bf 4.} 
We consider atmospheric movements associated with heat processes and focus on a closed system of dry air masses.
The next remarks refer only to processes in which the condensation has no part.
As a closed system, one can see the whole air bowl.
In the assumption of such a system there is no restriction of the task.
Sometimes it is convenient to choose a smaller air mass.
We want to think them open from the ground and rigid vertical walls open upwards.

What amount of kinetic energy can be generated in the system from known initial conditions with given supply of heat?

$q$ is the heat supplied to the air mass from the outside (which transitions into the boundary or is radiated into the room has already been subtracted).

A part of $q$ remains as heat and increases the internal energy of the air around $\delta I$.

A second part does work against the gravity, the potential energy of the situation grows by $\delta P$.

We do not consider other than mechanical and thermal forms of energy.
Therefore, the increase of the kinetic energy $\delta K$ can be written by the equation:
\begin{equation}
   \hspace{6cm}
    \delta K \;  =  \;  q  \: -  \:  \delta ( P + I )    
  \hspace{3cm} .................. \hspace{1cm}  (E)  \nonumber
\end{equation}

Instead of this form, other treatise on the energy of the storms has introduced another form which is, in some cases, advantageous:
\begin{equation}
   \delta K \: + \: (K) \;  =  \; (Q)  \: -  \:  \delta ( P + I )    \: .
   \nonumber
\end{equation}
$(K)$ is the loss of kinetic energy by friction and (Q) is the heat input, including frictional heat.
A part of $(K)$ can remain as heat in the air masses $r_e$, the other part $r_w$ lies in the boundary or is radiated.
If $(q)$ is the heat supplied to the air from outside, we have
\begin{equation}
   (Q) \;  =  \; (q)  \: + \:  r_e  \;  =  \; (q)  \: + \: (K)  \: -  \: r_w 
   \nonumber
\end{equation}
and for the $q$ introduced above
\begin{equation}
   q \;  =  \; (Q)  \: - \:  (K)  \;  =  \; (q)   \: -  \: r_w     \: .
   \nonumber
\end{equation}

Equation (E) states that kinetic energy does not necessarily arise at the expense of the simultaneously supplied heat.
$K$ can grow without heat supply when $P + I$ decreases.
In this case, the sum of the potential energy of the position and the internal energy is to be viewed as the total potential energy.
The internal energy $I$ appears as the potential of the internal forces, which is the (normal) compressive force alone, since we introduce for air the state equation of the ideal gas.
If any small virtual shift $P + I$ decreases, the system is not in equilibrium (or ``unstable'').
The state is stable when $P + I$ has a minimum value.

(1) If the layer of a stationary column of air is heated until buoyancy and kinetic energy are produced, then one can ask which portion of the heat supplied has passed into that form of energy?

(2) If the layer is cooled, there can also be {\it lively power\/}.
But it would not make sense to ask which part of the heat withdrawn has turned into kinetic energy.
This arises from the supply of potential energy.
The sum $P+I$ had a certain value before cooling, which was a minimum in relation to all possible small displacements.
After cooling, the sum $P + I$ has a smaller value which, however, is not a minimum value for the new state.

To answer the question in (1), we must know $(P + I)_1$ before heating, $(P + I)_2 = q + (P + I)_1$ after heating before the start of the movement, and $(P + I)_3$ at the time when the kinetic energy is the greatest value, leading to
\begin{equation}
   \delta K \;  =  \; (P + I)_2  \: - \:  (P + I)_3  
               \;  =  \; q   \: -  \:  
                        \left[ \:  (P + I)_2  \: - \:  (P + I)_1 \: \right]    \: .
   \nonumber
\end{equation}

Here $q$ and the subtracted term are positive.
In (2) they are both negative, and the absolute value of the subtracted term is larger.

If air movements are caused by heat processes, but are not sustained by such processes, the condition which occurs after heating or cooling is selected as the initial state.
From then on, the kinetic energy is fed only from the entire potential:
\begin{equation}
   \hspace{5cm}
    \delta K \;  =  \;   -  \:  \delta ( P + I )    
    \hspace{5mm} \mbox{for} \hspace{5mm}
    q \: = \: 0 \: .
  \hspace{1cm} .................. \hspace{1cm}  (E*)  \nonumber
\end{equation}

In order to make use of the energy equation, one has to calculate $P$ and $I$.
If $T$ is the absolute temperature of the mass part $dm$ at the height $z$\footnote{Further, $C_p$ and $C_p$ are the specific heat of the air at constant volume and constant pressure, $C_p - C_v = R$ is a constant in the state equation $p = R \: T \: \mu$, $p$ the pressure, $\mu$ the density; $g$ the Gravity acceleration; and $C_0$...$C_2$ are Constants in the difference terms.}, 
one has
\begin{equation}
    \mbox{a)} \hspace{2mm}  I \;  =  \;   C_v \: \int T \: dm  \: + \: C_1  \; ; 
    \hspace{8mm}
    \mbox{b)} \hspace{2mm}  P \;  =  \;   g \: \int z \: dm    \: .
    \nonumber
\end{equation}

These terms apply generally. 
The second term can be made into a convenient form for the calculations if the following assumptions are made for the initial and final states:

I. The pressure in each location is equal to the weight of the overlying unit column.

II. The air mass consists of two parts separated by a displaceable level surface, the upper mass acting only as a constant weight punch and the lower mass and stamp together form the closed system.

The assumption II does not seem to imply a significant restriction of the task, since the height of the limiting level can be selected as desired.
Then it will be
\begin{eqnarray}
   & & \mbox{b*)} \hspace{2mm}  P \;  =  \;   R \: \int T \: dm  \: + \: C_2 \: , 
    \hspace{8mm}
    \nonumber \\
   & & \mbox{c)} \hspace{2mm}  P  \: + \: I  \;  =  \;    C_p \: \int T \: dm  \: + \: C_0   \: .
    \nonumber
\end{eqnarray}

Further, if $T$ and $T'$ are valid for the initial and final states of the same air particle $dm$, the energy equation is:
\begin{eqnarray}
\hspace{3cm}
   & & \delta K \;  =  \;  q \: - \:  C_p \: \int (T' \: - \: T) \: dm  
    \hspace{2.9cm} .................. \hspace{1cm}  (E_1)  \nonumber
    \nonumber \\
\hspace{3cm}
   & & \delta K \;  =  \;   C_p \: \int (T \: - \: T') \: dm
    \hspace{5mm} \mbox{for} \hspace{5mm}
    q \: = \: 0 \: .
    \hspace{1cm} .................. \hspace{1cm}  (E_1^{\ast})  \nonumber
    \nonumber
\end{eqnarray}
These integrals are to extend over the mass under the punch.

Change of centre of gravity of an air column is supplied by heat.
These equations are to be examined first by a static survey.
Some layers of a still air column, more generally a still mass in the closed system, are supplied with heat. 
The layers can remain in the same sequence, or they can be pushed together, and the kinetic energy is finally consumed by friction.
According to ($E_1$) with $\delta K = 0$ and with b*) for a stamp zero, the height $\delta \zeta$ for raising the centre of gravity for the whole mass $M$ is given by
\begin{equation}
   g \: M \: \delta \zeta \;  =  \;   \delta P  \;  =  \;  R \: \int (T' \: - \: T) \: dm   \; ; 
   \hspace{5mm}
   q \;  =  \;   C_p \: \int (T' \: - \: T) \: dm   \; ; 
   \hspace{5mm}
   \delta \zeta \;  =  \;  \frac{R}{C_p} \: \frac{q}{g \: M}    \; .
   \hspace{5mm}
   \nonumber
\end{equation}
For a unit column, $g \: M$ is equal to the ground pressure $p_0$ and the centre of gravity is raised by the amount
\begin{equation}
   \delta \zeta \;  =  \;  \frac{R}{C_p} \: \frac{q}{p_0}    \; ,
   \hspace{5mm}
   \nonumber
\end{equation}
irrespective of the places where heat has been supplied, it is proportional only to the whole feed.

The position of the centre of gravity is not changed by internal heat exchange. (See Ritter, Applications of the Mech. Heat Method to Cosmol. Probl., p.46-49. Leipzig, 1862.)

\vspace{3mm}
 $\bullet$ {\bf 5.} {\it Available kinetic energy\/} of the air mass in a closed system. Example of the two chambers.
 
The air mass is now in a certain state. 
We ask: what is the greatest amount of kinetic energy that it can leave to itself? 
The initial value $(P + I)_a$ of the total potential energy is given.
The question can easily be answered if the smallest possible final value $(P + I)_{\rm min}$ can be specified.

Let us imagine that the mass is transferred adiabatically into a new state in all parts, so that now every level becomes an isothermal surface, and the layers are ordered according to the rising potential temperature.
Then the mass, if it had no kinetic energy, could remain in stable equilibrium.
This is the only arrangement of this kind, i.e., the position of the smallest potential energy $(P + I)_{\rm min}$ in adiabatic condition.

It can be shown that this is the smallest value which can be obtained without heat expansion.
We bring the system into the stratification of stable equilibrium, first (A) in the smallest parts isentropic, then (B) after the small masses $m_1$ and $m_2$ have mixed the different potential temperatures during the operation.
In (A), $m_1$ and $m_2$ form separate layers, and between them there are others of intermediate entropies.
In order to obtain state (B) from (A), the two layers of the gravitational force should be displaced to the position where $m_1 + m_2$ lies in (B).
One would have to work to produce (B) from (A).
Accordingly, $(P + I)_a$ is less than $(P + I)_b$ and generally $(P + I)_{\rm min}$ is the value obtained for stable stratification with isentropic layer change of each mass part, which is followed by a layering after mixing.

One has the greatest value of the increase of kinetic energy:
\begin{equation}
   (\delta K)_{\rm max} \;  =  \;  (P + I)_a  \: - \: (P + I)_{\rm min}  \; ,
   \hspace{5mm}
   \nonumber
\end{equation}

If the assumptions I and II are true, the last term on the right is easy to compute.
For every mass $dm$ in the initial state $p$ and $T$ are known.
The  weight $p '$ of the plunger plus that of the higher potential temperature layers for the surface unit is also used for the state of the smallest potential energy, so that from the equation of the isentropic state change the new temperature $T'$ can also be applied to equation ($E_1^{\ast}$).

Let us now consider the example of two rooms.
Let us imagine that each is a few kilometres high, many hundreds of miles long and wide, and that both bear a movable stamp instead of the ceiling.
Here we have the task of solving the first of the storm theories discussed above.

Initial condition: the air mass is bounded by flat ground and vertical walls.
A colder part ``1'' above the base $B_1$, a warmer part ``2'' over $B_2$, are separated.
In each part the temperature is a function of the height alone, and the stratification is stable.
The potential temperature of the highest layer of 1 is lower than that of 2 on the ground.
The average temperatures are $T_1$ and $T_2$. 
Both masses reach the level $h$, with above them a constant pressure level. 

The masses start to move. 
We obtain $ (\delta K)_{\rm max}$ when the colder mass 1 has spread over the entire bottom surface $B = B_1 + B_2$, and the layers in 1 and 2 have the same sequence as in the initial state.
If $ (\delta K)_{\rm max}$ were evenly distributed over all mass parts, each would have the velocity:
\begin{equation}
   V \;  =  \;  \sqrt{\: \frac{B_1 \: B_2}{B^2} \: \tau \: g \: h \;}
    \hspace{5mm} \mbox{with} \hspace{5mm}
    \tau \: = \: \frac{\; T_2 \: - \: T_1 \;}{\sqrt{\: T_1^{\phantom{V}} T_2 \:}}
     \: .
    \nonumber
\end{equation}
The expression has the greatest value for given $B$ if $B_1 = B_2$, leading to
\begin{equation}
  \hspace{6cm}
   V \;  =  \;  \frac{1}{2} \; \sqrt{\: \tau \: g \: h \;}
     \: .
    \hspace{3cm} .................. \hspace{1cm}  (1)  \nonumber
    \nonumber
\end{equation}
This is the velocity of a vacuum-free body, which has decreased by $\tau\:h/8$.

With $T_1 = 273$~K, $T_2 = 283$~K and $h = 3000$~m, then $V = 16$~m/s, and for $h = 6000$~m then $V = 23$~m/s.
With lower mean temperatures but with the same difference, $V$ becomes somewhat larger.

Air masses, which have temperature differences of up to a few kilometres, contain a supply of potential energy sufficient to generate storms in the whole field.
In reality, the kinetic energy is not evenly distributed into the reacting masses. Storm will prevail in a part of the area long before $ (\delta K)_{\rm max}$ is reached, even if part of the {\it living force\/} is consumed by friction, and even if the application is not strictly isentropic.

In (1) it is assumed that the masses were initially separated by a vertical wall.
If the wall forms the angle $\delta$ with the vertical, the volume of the chambers being the same, if the cold mass is located at the bottom in the acute wedge, and if $l$ is the length of the whole trough of the same width, then
\begin{equation}
   V \;  =  \;  \frac{1}{2} \; \sqrt{\: \tau \: g \: h \;}
               \; \sqrt{\: 1 \: - \: \frac{1}{3} \: \frac{h}{l}  \: \tan(\delta) \;}
    \hspace{5mm} \mbox{for} \hspace{5mm}
      \tan(\delta)  \: < \: \frac{l}{h}
     \: .
    \nonumber
\end{equation}

The calculated example can be regarded as a model of the processes in gusts.
In other storms, there are usually constant temperature distributions.
For these, too, the calculation can easily be carried out.

In  equation (1), it first becomes clear that $V$ is independent of the kind of the gas.
The constants $R$ and $C_p$ do not impact.
The same $V$ is thus obtained at the same temperature in hydrogen as in air. 
This is not very accurate.
Equation (1) is thus only an approximation, but a very useful one.

The same $V$ is obtained if, in place of the air masses of different medium temperatures, there are different liquids of unequal density, and if $\tau$ denotes the relative difference in density (quotient of the density difference by the mean density; namely $(\rho_2-\rho_1)/\sqrt{\rho_2 \: \rho_1\:}\:$?). 
The kinetic energy produced during the transfer is essentially an effect of the gravitational force, although the internal energy also decreases in the gas due to the change of the energy of position (namely: $g \: z$?).

With the assumption II, the ratio $\delta I / \delta P$ is equal to $C_v / R$  [according to the equations a) and b*)], leading to $1 / 0.41$ for the air.
The calculation can also be carried out without this assumption.
The masses can be allowed to remain constant in volume, exactly according to the example of the two rooms, or to remain the pressure at the confining level, so as to produce no appreciable amount of kinetic energy.
Then the equilibrium (1) is again obtained, but the ratio $\delta I / \delta P$ is quite different than before.

In these systems, the generation of a large amount of kinetic energy depends on the fact that masses from the colder part sink by a few kilometres more than those of the warmer ones. 
The vertical displacements are therefore small compared with the horizontal length, and the inclination of the orbits against the levelling surfaces can remain far smaller than one arc degree.

Horizontal pressure differentials must consist of masses of different temperatures at the same level.

The horizontal pressure gradient can grow in a part of the area during the movement.
There is nothing in the energy calculation, nor in the effects of the distracting force, for these do not contribute to the kinetic energy.
Pressure gradients, and trajectory forms, with which the mechanics of meteorologists are almost exclusively interested in, whose importance for dynamic tasks is evident, can be ignored if one looks only at the {\it available kinetic energy\/} of a system.

However, the work of the compressive forces is also included in our calculation, although in a hidden form as part of $\delta I$. 
This can be seen in the following.

\vspace{3mm}
$\bullet$ {\bf 6.} {\it Available kinetic energy\/} of a horizontal air disk.

Let us imagine a closed horizontal air-disk, so thin that one can ignore the vertical pressure gradient, and that the whole mass is shifted by an imperceptibly small $\delta P$. 
The initial state is given by a certain pressure distribution $p$.
The mass comes into motion and gains the greatest kinetic energy without friction, if the pressure $p'$ is equal in the whole disk.
If the temperature of the mass $dm$ in the initial state is $T$, and if the temperature after the equalizing in pressure is  $T'$, we have:
\begin{equation}
  \hspace{7mm}
   \delta K \;  =  \;  - \: \delta I \; = \; 
     C_v \int (T - T') \: dm 
    \hspace{5mm} \mbox{for} \hspace{3mm}
       \delta P \; = \; 0
    \hspace{5mm} \mbox{and} \hspace{3mm}
       q \; = \; 0
    \hspace{5mm} .................. \hspace{1cm}  (E^{\star \star})  \nonumber
    \nonumber
\end{equation}

If the mass from the state $p'$ is to be transferred adiabatically into the pressure distribution $p$, one has to perform the work $W$ against the compressive force.
$W$ can be calculated \footnote{Working value of an air pressure distribution. Anniversary volume of C.-A. F. Met. Wiener Denkschriften (1901) 73, 330, (ibl) with the modified name.} by
\begin{equation}
   W \;  =  \;
     \frac{C_v}{R} \: \int ( \: p \: - \: p' \: ) \: dk 
     \; = \; 
     C_v \: \int (\: T \: \mu  \: - \: T' \: \mu' \:) \: dk 
     \; = \; 
     C_v \: \int (\: T   \: - \: T'  \:) \: dm
  \: ,  \nonumber
\end{equation}
where $dk$ is the volume element.

If the mass comes by itself from the state $p$ to $p'$, the compressive forces are subject to a compression.
It is identical with the decrease of the internal energy.
One can say, therefore, that the kinetic energy arises from the work of the compressive forces or from the internal energy.
The first formulation corresponds to the principle of {\it living force\/}, the other to the energy principle.

This does not apply to the air disc alone.
In any closed system, the work of the compressive forces in transition from one state to another is equal to the decrease of the internal energy when no heat is supplied from the outside.
The internal energy $I$, as already mentioned above, is to be seen as a potential of the compressive forces in the closed adiabatic system.

Now we assume the following initial state in the disk: 
a partial volume $k_1$ has the pressure $p_1$, the other $k_2$ the pressure $p_2$.
The greatest kinetic energy that can be generated in this system is calculated from (E **) with $T'$ corresponding to the balanced pressure $p'$.
If this $\delta K$ is distributed in the whole mass, and if $k_1 = k_2$, then each particle has the velocity
\begin{equation}
   (V) \;  =  \; \frac{1}{2} \; \;  \sqrt{\: \frac{C_v}{C_p} \: R \: T \;}
                 \; \; \frac{\: p_1 \: - \: p_2 \:}{p_1}
     \: ,
    \nonumber
\end{equation}
where $T$ is the average temperature of the mass.
For  $p_1 = 765$~mm-Hg, $p_2 = 755$~mm-Hg and $T = 273$~K,  then $(V) = 1.55$~m/s.

We compare the velocity $(V)$ of the disk with the velocity $V$ of the two-chamber system of considerable height, when $p_1$ and $p_2$ have the same values in the first chamber, and with the initial pressures $p_{01}$ and $p_{02}$ at the bottom of the second chamber.
We use an already calculated example (En. D. St., p.14).

The pressures are $p_{01} =759.2$~mm-Hg, $p_{02} = 763.5$~mm-Hg, $p_h=510$~mm-Hg.
These branches belong to two chambers of $3000$~m in height, whose average temperatures are $T_1 = 257.9$~K and  $T_2 = 262.9$~K.
From (1), one obtains $V = 12.2$~m/s.
With $p_1 = 759.2$~mm-Hg, $p_2 = 753.5$~mm-Hg and $T = 260.4$~K, from (2) one obtains $(V) = 0.89$~m/s.

When the partition of $k_1$ and $k_2$ is pulled away, the velocity of the parts near the wall can be much greater than the mean $(V)$.
As soon as the motion is distributed over the whole mass, the velocity is everywhere of the order of magnitude of the calculated $(V)$.
A tenfold larger could be produced only in the hundredth part of the mass.
But also the storm during the fraction of a second does not apply if the initial pressure distribution is continuous.

We now set for the initial state in the disc: $p = p' \: (1 + \epsilon)$, where $\epsilon$ is partly positive, partly negative, everywhere continuous, and $p'$ as above is  the pressure after adiabatic adjustment.
The supplied work of this system \footnote{Working value of an air pressure distribution p. 331, equation (Ib *).} is:
\begin{equation}
  W  \;  =  \; 
     \frac{C_v}{C_p} \; p' 
      \int \: \frac{\epsilon^2}{2} \:  dk
 \; = \; 
     \frac{C_v}{C_p} \; p' \;   \frac{k}{2} \;
      [ \: \epsilon^2 \: ]
     \: ,
    \nonumber
\end{equation}
where $[ \: \epsilon^2 \: ]$  is the average of $\epsilon^2$ over the whole slice.
The mass of air is $k \: p' / (R \: T)$, which is the average vital force of the mass unit, which can arise from W
\begin{equation}
  \hspace{7mm}
   \frac{(V)^2}{2}
   \; =  \;
     \frac{C_v}{C_p} \; R \; T \; 
     \frac{[ \: \epsilon^2 \: ]}{2} \; ,
    \hspace{5mm} \mbox{and approximatively} \hspace{3mm}
   (V) \; = \; 
        \left( \;
   \sqrt{\:  \frac{C_v}{C_p} \: R \: T \;}
       \; \right)
   \:  [ \: \epsilon \: ]
   \; ,
    \nonumber
\end{equation}
where $[ \: \epsilon \: ]$ replaces $\sqrt{\: [ \: \epsilon^2 \: ] \:}$.

The root expression in $(V)$ is $1/1.41$ of the sound velocity.
With $T = 273$~K, $(V) = 236$~m/s.

The final approach will be used in Figure.~3 to conclude that unsuccessful gradients do not generate a storm.

\section{\underline{Calculation of the kinetic energy, which can result from}\\
\underline{the heat of condensation}.\protect\footnote{This section contains the supplement promised in the treatise on the energy of storms S.4}}
\label{section_III}

\noindent \hspace{3 mm} $\bullet$ {\bf 7.}
In the systems we will now introduce, we distinguish a mass 1, which remains in the dry state, for which we apply dry air, and a saturated mass 2, the vapor of which partly condenses.
For a particle of 1, which passes adiabatically from the state $p_1$, $T_1$ to another $p'_1$, $T'_1$:
\begin{equation}
  T'_1  \;  =  \; T_1 \; 
        \left( \;
           \frac{p'_1}{p_1}
       \; \right)^{\varkappa}
    \hspace{5mm} \mbox{with} \hspace{3mm}
    \varkappa \; = \; \frac{R}{C_p} \; = \; \frac{0.41}{1.41}
     \: .
    \nonumber
\end{equation}
For a particle of 2, a similar equation can be used according to the procedure of Heye:
\begin{equation}
\hspace{60mm}
  T'_2  \;  =  \; T_2 \; 
        \left( \;
           \frac{p'_2}{p_2}
       \; \right)^{\lambda} \: ,
    \hspace{30mm} .................. \hspace{1cm}  (1)  \nonumber     
    \nonumber
\end{equation}
where $\lambda$ is positive and smaller than $\varkappa$ and is a function of the three quantities $T_2$, $p_2$, $p'_2$.
The values of $\lambda$ vary in the atmosphere between $\varkappa / 3$ and $\varkappa$.
This makes the calculations for moist air more extensive than for dry ones.
However, we do not need any exact figures here, we just want to get an overview of the conditions under which the condensation heat contributes to the kinetic energy and estimate the amount in individual cases.
Therefore, the variability of $\lambda$ is of little importance for our task. 
We put:
\begin{equation}
    \lambda \; =  \; \mbox{constant} 
    \nonumber
\end{equation}
and will have to choose their value in every calculations.

Now let us imagine that the gas 2 is dry air, which is supplied with just as much heat as to satisfy equation (1) upon expansion ($p'_2 <p_2$).
The condensation heat for the mass unit is
\begin{equation}
    \hspace{50mm}
  Q  \;  =  \;
          \left(
           \frac{\varkappa \: - \: \lambda}{\lambda}
          \right)
           \; C_p \;  \left(  T_2 \: - \: T'_2 \right) \: .
    \hspace{20mm} .................. \hspace{1cm}  (2)  \nonumber     
    \nonumber
\end{equation}

Steam containing air has lower density than dry air at the same temperature and pressure.
We take this fact into account by defining
\begin{equation}
    T_2, T'_2 \; =  \; \mbox{the virtual temperatures} 
    \nonumber
\end{equation}
as the temperatures of dry air of the same density at $p_2$ at $p'_2$, with $\lambda$ defined accordingly.

If $(T_2)$ is the observed temperature, if  $T_2$ is the virtual temperature (Guldberg and Mohn), if $p$ is the pressure and $p_{\beta}$ is the partial pressure of the steam, so the relationship exists
\begin{equation}
    T_2 \; =  \; 
      \frac{(T_2)}{ 1 \: - \: 0.3767 \: p_{\beta}/p} \: .
    \nonumber
\end{equation}
The air which only expands with the addition of heat to the equations (1) and (2), and which takes the place of the steam-saturated air, is sometimes called the ``fictive gas''.
In a column of this gas, $b$ is the gradient of temperature:
\begin{equation}
    \hspace{55mm}
    b \; =  \; 
      - \: \frac{dT}{dz} 
     \; =  \; 
     \frac{\lambda}{\varkappa} \;
     \frac{g}{C_p} 
     \; =  \; 
     \lambda \;
     \frac{g}{R}    \: .
    \nonumber
    \hspace{20mm} .................. \hspace{1cm}  (3)  \nonumber     
\end{equation}

For a closed system, consisting of the masses 1, 2 and the stamp, the energy equation ($E_1$) holds:
\begin{equation}
   \delta K \;  =  \;  
   q 
   \: - \:  C_p \: \int (T'_1 \: - \: T_1) \: dm_1  
   \: - \:  C_p \: \int (T'_2 \: - \: T_2) \: dm_2  
    \: ,
    \nonumber
\end{equation}
wherein the heat input of the equation (2) is determined by
\begin{equation}
\hspace{50mm}
  q  \;  =  \;
          \left(
           \frac{\varkappa \: - \: \lambda}{\lambda}
          \right)
           \; C_p \;  \int (  T_2 \: - \: T'_2 ) \: dm_2 \:  . 
    \hspace{10mm} .................. \hspace{1cm}  (4)  \nonumber     
    \nonumber
\end{equation}
This means
\begin{equation}
\hspace{28mm}
  \delta K  \;  =  \;
    C_p \: \int (T_1 \: - \: T'_1) \: dm_1  
   \; + \;
    \frac{\varkappa}{\lambda} \;
    C_p \: \int (T_2 \: - \: T'_2) \: dm_2  \:  . 
    \hspace{10mm} .................. \hspace{1cm}  (5)  \nonumber     
    \nonumber
\end{equation}

\begin{figure}[hbt]
\centering
\includegraphics[width=0.5\linewidth,angle=0,clip=true]{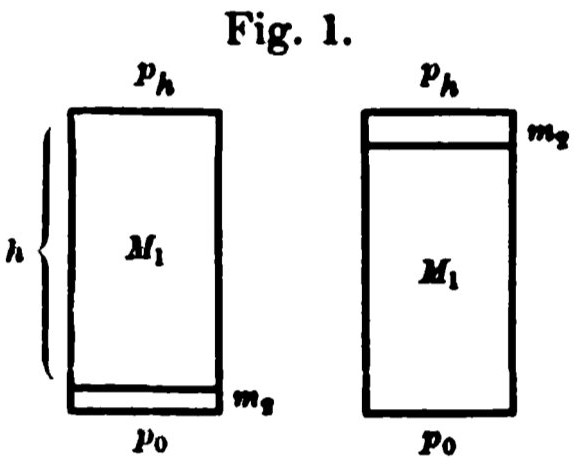}
\label{fig_1}
\end{figure}

\vspace{3mm}
$\bullet$ {\bf 8.} We apply these equations first to the following problem (Fig.1).

In the initial unit column (on the left) there is a very thin layer of steam-saturated air of mass $m_2$ (in the calculation: ``fictive gas''), with a pressure $p_0$ below.
Above is $M_1$, a column of dry air of height $h$, bounded by the stamp, whose weight is $p_h$.
In the final state (on the right): $m_2$ has ascended and has spread between $M_1$ and the punch.
What amount of kinetic energy is generated, which part of the condensation heat remains as heat in the system?

We have the following relations for the final temperatures of $m_2$ and all the layers of $M_1$ (the sequence of layers is retained here):
\begin{equation}
  T'_2  \;  =  \; T_2 \;
       \left( \;
           \frac{p_h}{p_0}
       \; \right)^{\lambda} \: , \; \; \;
  T'_1  \;  =  \; T_1 \:
       \left( \; 1 \: + \:
           \frac{g \: m_2}{p_1}
       \; \right)^{\varkappa}
    \:  .
    \nonumber
\end{equation}
With the assumption that $g \: m_2$ is small against $p_h$, it follows \footnote{As in the energy of the storms, p.8.}
\begin{equation}
  C_p \: \int (\: T_1 \: - \: T'_1\: ) \: dm_2  \;  =  \;  - \: g \: h \: m_2 \;
    \:  .
    \nonumber
\end{equation}
In order to calculate $T_2$, if we use different values for $h$, one must know the temperature layer in $M_1$.
We take it varying linearly and call the temperature gradient $a$, leading to:
\begin{eqnarray}
 T_1  &  =  &  T_{01} \: - \: a \: z \; , \; \;
     \hspace{5mm} \mbox{therefore} \hspace{5mm}
 \frac{p_h}{p_0}  \;  =  \;
       \left( \; 1 \: - \:
           \frac{a \: h}{T_{01}}
       \; \right)^{g/(a \, R)}
    \:  ,
    \nonumber \\
    T_2  \: - \: T'_2 &  =  & 
    \lambda \; \;  \frac{g \: h}{R} \; \;
    \left\{ \:
      1 \: + \: \frac{a-b}{2} \: \frac{h}{T_{01}} \: + \: \ldots \;
    \: \right\} \:
    \frac{T_2}{T_{01}}
    \:  .
    \nonumber 
\end{eqnarray}
In order to replace $\varkappa$ and $\lambda$ in the formulas by other clear quantities, we introduce a temperature gradient of the indifferent state of dry air:
\begin{equation}
 A  \;  =  \;  \frac{g}{C_p} 
    \; = \;
    \varkappa \; \frac{g}{R} \; , \; \; \; \; \; \; 
 \frac{\varkappa}{\lambda}
    \; = \;
 \frac{A}{b}
    \:  .
    \nonumber 
\end{equation}

In Equations (4) and (5), $T_2$ and $T'_2$ are the same for all parts of the small layer $m_2$.
When $a \: h$ or $T_{01} - T_{h1}$ is smaller than $T_{01}$, the approximate values are:
\begin{eqnarray}
\hspace{40mm}
 q  &  =  &  m_2 \; \;  \frac{a \: - \: b}{A} \; \; \frac{T_2}{T_{01}} \; \;  g \: h
    \:  ,
    \hspace{35mm} .................. \hspace{1cm}  (4_1)  \nonumber     
    \nonumber \\
\hspace{40mm}
 \delta K   &  =  &  m_2 \;
    \left( \:
         \frac{T_2 \: - \: T_{01}}{T_{01}}
         \: + \:
         \frac{a-b}{2} \: \frac{h}{T_{01}}  \: \frac{T_2}{T_{01}}
    \: \right)  \; g \: h
    \:  .
    \hspace{2mm} .................. \hspace{1cm}  (5_1)  \nonumber     
    \nonumber 
\end{eqnarray}
These equations also apply when $m_2$ is dry air.
Then one has $b = A$ and $q = 0$.
The increase in kinetic energy $\delta K$ can be positive if $T_2 >  T_{01}$, namely if the column rises in warmer dry mass  than with the temperature gradient $a$.

\vspace{3mm}
$\bullet$ {\bf 9.} Let us set
\begin{equation}
 T_2  \;  =  \;  T_{01} \; , \; \; \; \; \; \; 
 a \; > \; b 
    \:  ,
    \nonumber 
\end{equation}
where an unstable initial state has been assumed as if a vapor-saturated layer is under dry air of equal density, whose vertical temperature gradient is greater than that of the indifferent state of the steam-saturated mass.
The mass $m_2$ can then rise with the buoyancy:
\begin{equation}
 q  \;  =  \; m_2 \; \; \frac{A \: - \: b}{A} \; \;  g \: h
     \; , 
\hspace{2mm} ...... \hspace{2mm}  (4_2) 
\hspace{20mm}
 \delta K  \; = \; m_2 \; \frac{a \: - \: b}{a} \; \;  \frac{h}{T_{01}} \; \; g \: h
    \:  .
\hspace{2mm} ...... \hspace{2mm}  (5_2)   
    \nonumber 
\end{equation}

Example: a mass $m_2$ of saturated air at $20$~C, whose absolute virtual temperature is $T_2 = 295.5$K.
$M_1$ is a column of dry air with $a = 0.007$~C/m, $h = 1000$~m, $T_{01} = 296.5$~K, $T_{h1} = 288.5$~K.
Calculations are made with $p_0 = 760$ and $p_h = 676.1$~mm-Hg.

The mass $m_2$ rises and comes under the pressure $p_h$.
For saturated air, the final temperature is $15.5$~C.
The associated virtual temperature is $T'_2 = 290.6$~K.
From (1) and (3), the values of $\lambda$ and $b$ fitted for our case can be found by substituting $p_0$ and $p_h$ in (1), instead of $p_2$ and $p'_2$, leading to:
\begin{equation}
 \lambda  \;  =  \;  0.14294 
 \; , \; \; \; \; \; \; 
 b \; = \; 0.00488 \; \mbox{C/m} 
    \:  .
  \nonumber 
\end{equation}

With $A = 0.00993$~C/m and $m_2 = 1$~kg, one obtains from ($4_2$) and ($5_2$):
\begin{equation}
 q  \;  =  \;  4990 \; \mbox{kg.(m/s)${}^2$} \; = \; 12 \: \mbox{Cal}
 \; , \; \; \; \; \; \; 
 \delta K  \; = \; 35.2 \; \mbox{kg.(m/s)${}^2$} 
 \:  .
  \nonumber 
\end{equation}

The fraction of heat supplied (the condensation heat), which passes into {\it living force\/}, is equal to $\delta K / q = 1/140$.

If all the kinetic energy were concentrated in the ascending mass, this would get a velocity of $8$~m/s in $1000$~m.

The term $\lambda$ of the fictional gas is determined so that to satisfy the virtual initial and final temperature of the steam-saturated air.
In order to know the term $q$, we need a direct calculation of the condensation heat of this air.
Since $1$~kg of saturated at $20$~C contains $0.0146$~kg of steam, and since the final state at $15.5$~C contains $0.0124$~kg of steam, the mass of condensed water is $0.0022$~kg.
At the average temperature of $18$~C, the condensation heat is thus $596$~Cal/kg and $0.0022 \times 596 = 1.3$~Cal.

According to the equation ($5_2$), if $b$ is constant, the value of $\delta K$  would increase with height proportionally with $h^2$.
But $b$ is an average value to be determined each time for the chosen $h$, and which grows with $h$.
If $a <A$, the factor $a-b$ becomes smaller and $\delta K$ has a maximum value for a certain height.
Only for small altitudes $\delta K/q$ is nearly proportional to the height, around $m_2$.

We are still looking at the case
\begin{equation}
 T_2  \;  <  \;  T_{01} \; , \; \; \; \; \; \; 
 a \; > \; b 
    \:  , \; \; \; \; \; \; 
 \mbox{Equations ($4_1$) and ($5_1$)}
 \: .
    \nonumber 
\end{equation}

If the mass $m_2$, which is saturated thereon, has a smaller virtual temperature (greater density), the equilibrium is initially stable.
If, however, $m_2$ is raised by a small impulse, it can rise further with buoyancy. 

At the expense of work (from the kinetic energy supply on the system), the particles are brought up to a height ($h$) in which  $\delta K$ has a minimum value in equation ($5_1$) :
\begin{equation}
 (h)  \;  =  \;  
  \frac{T_{01} \: - \: T_2}{a \: - \: b} \; \;
  \frac{T_{01}}{T_2} 
 \: .
    \nonumber 
\end{equation}

From this point,  $m_2$ increases with the lift till the height
\begin{equation}
 [h]  \;  =  \;  2 \: (h) 
 \: 
    \nonumber 
\end{equation}
is reached, for which $\delta K = 0$.
Only when further ascending does $K$ increase.

Example: for $T_2 = 295.5$~K (average air at $20$~C),  $T_{01} = 296.5$~K, $a = 0.007$ and $b = 0.00488$, then $(h) = 470$~m and $[h] = 940$~m.

\begin{figure}[hbt]
\centering
\includegraphics[width=0.7\linewidth,angle=0,clip=true]{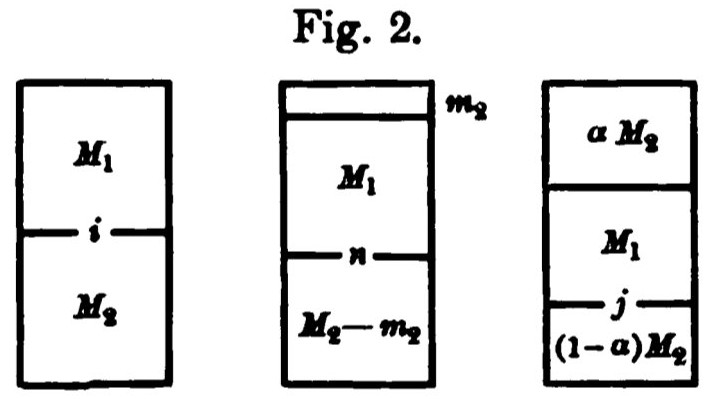}
\label{fig_2}
\end{figure}

\vspace{3mm}
$\bullet$ {\bf 10.} In the unit column (see Fig.2), we now place the mass $M_2$ of steam-saturated air under the dry column $M_1$ in the indifferent state (or the fictive gas), with a piston over $M_1$.
With the index $i$ we characterize the initial values at the boundary of $M_1$ and $M_2$.
The initial temperature distribution is:
\begin{eqnarray}
 T_{i1}  \; - \; a \; (z \: - \: z_i )  \; & \mbox{in} & \; M_1 \; ,
    \nonumber \\
 T_{i2}  \; + \; b \; (z_i \: - \: z )  \; & \mbox{in} & \; M_2 \; .
    \nonumber 
\end{eqnarray}

From $M_2$, the small boundary layer $m_2$ at initial height $\zeta$ rises over $M_1$, under the stamp $p_h$.
The pressure at the new boundary surface is $p_n = p_i + g \: m_2$.
All values on this surface are given the index $n$.
For a sufficiently small $\zeta$ we have
\begin{equation}
 m_2  \;  =  \;  
    \mu_{i2} \;  \zeta
 \;  =  \;
    \frac{p_i}{R \: T_{i2}} \;  \zeta
 \; ; \; \; \;  \; \; \; \; \; \;
 p_n  \;  =  \;  p_i
    \left( \:
      1 \: + \: \frac{g \: \zeta}{R \: T_{i2}}
    \: \right) \:
    \;  =  \;  p_i
    \left( \:
      1 \: + \: \frac{A}{\varkappa} \: \frac{\zeta}{T_{i2}}
    \: \right) \:
    \:  .
    \nonumber 
\end{equation}
Moreover
\begin{equation}
 T_{n1}  \;  =  \;  
    T_{i1} \;  
           \left( \;
           \frac{p_n}{p_i}
       \; \right)^{\varkappa} 
 \;  =  \;
    T_{i1}
    \left( \:
      1 \: + \: \frac{A \: \zeta}{T_{i2}}
    \: \right) \:
    \: , \; \; \; \; \; \; \; \; \;
 T_{n2}  \;  =  \;   T_{i2} \; + \; b \; \zeta 
   \:  .
    \nonumber 
\end{equation}

When $T_{i2} = T_{i1}$ after the small mass $m_2$ has risen from the layer $\zeta$ and over $M_1$, the new boundary surface value  $T_{n1}$ becomes larger than $T_{n2}$:
\begin{equation}
 T_{n1}  \; - \;  T_{n2}  \; = \; ( \: A \: - \: b \:) \; \zeta \; .
    \nonumber 
\end{equation}

The initially unstable state on the surface $i$ has passed into a stable one at the boundary surface $n$.
A momentum is needed to lift the next layer of $M_2$ up to a height $(h)$ from which it can ascend.
However, it will still increase the kinetic energy of the system by rising, for the same mass by a somewhat smaller amount than the first $m_2$.
For each successive layer, the increment of $K$ is decreasing, and is finally zero when the boundary surface has decreased to $j$.
On the whole, $\alpha \: M_2$ has risen; Mass from the parts $(1-\alpha) \: M_2$ under $j$ could be passed through $M_1$ only with effort or with loss of kinetic energy of the system.

Example: the initial state of $M_2$ is a steam-saturated column of $2000$~m height in indifferent state, at the bottom of which $p_0=760$~mm-Hg and $t_{e2}=30$~C.

For these conditions, the following values of the temperature gradient at different heights are given in the tables of Neuhoff:
\begin{eqnarray}
   & & \hspace{2mm}  0 \:\mbox{m}     \hspace{18mm}       1000\:\mbox{m}     \hspace{15mm}     2000\:\mbox{m}  \nonumber \\
   - \: \frac{dt_2}{dz}  . . . . & &  
          0.0037\:\mbox{m}     \hspace{12mm}  0.0037     \hspace{15mm}   0.0038
          \hspace{5mm} \mbox{C/m}  \nonumber 
\end{eqnarray}
We put $0.0037$ everywhere, then we have the temperature $t_{i2} = 22.6$~C at $2000$~m.

Let us now calculate the absolute virtual temperatures at the boundary and the gradients related to them
\begin{equation}
 T_{e2} \; = \; 307.8 \: \mbox{K} \; ; \; \; \; 
 T_{i2} \; = \; 299.4 \: \mbox{K} \; ; \; \; \; 
 b \; = \;  \frac{8.4}{2000}  \; = \; 0.0042 \: \mbox{C/m} \; ; \; \; \; 
    \nonumber 
\end{equation}
(if we use the same $b$ in the calculation of $\delta K$, we will obtain a somewhat too great value, since further decrease in the steam content lead to smaller $a-b$, with $- \: dt_2 / dz$ and $b$ becoming larger.)

Initial state of $M_1$: altitude $2000$~m, $T_{i1} = T_{i2}$ (hence a state which is unstable at the boundary) and $a = A$.

This choice of $a$ is the most favourable when one wants to obtain large values of $\delta K$.
It also has the advantage that, in the case of conversions of parts of the mass $M_2$, the term $a$ remains unchanged (See Met, 1906, p. 242).

In order to arrive at the goal with the least possible effort of calculations, we suggest the following path:

(1) We let a kilogram of the uppermost layer of $M_2$ ascend, and use ($5_1$) or ($5_2$) to calculate the corresponding $\delta K$.

(2) We assume that almost the entire mass $M_2$ is already over $M_1$, and that only the last kilogram is still in a thin layer on the ground.
We then take $M_1$ above and the $\delta K$ are calculated for this.
If the kinetic energy of the system is positive, a rough approximation is found to be the product of $M_2$ by the mean value of the first and the last $\delta K$.
If, however, it is negative, as can be foreseen, the position of the layer (surface $j$) can be estimated from the values of the two calculated $\delta K$, whose ascent does not increase the kinetic energy of the system.

(3) Control of these approximations by computing $\delta K$ for the $1$~kg boundary area $j$.

In all these calculations equation ($5_1$) is to be applied with the following values:
\begin{eqnarray}
 m_2  \; = \; 1 \; \mbox{kg} \:  ,  
 &    &  
     \hspace{5mm}
 a  \; = \; A \; = \; 0.00993 \:  ,  
     \hspace{10mm}
 b \; = \; 0.0042 \; \mbox{C/m} \:  ,
    \nonumber 
\\
\mbox{and in the case (1)}  
 &    &  
     \hspace{5mm}
 T_2 \; = \; T_{01} \; = \; 299.4 \: \mbox{K} \:  ,  
     \hspace{5mm}
 h \; = \; 2000 \; \mbox{m} \:  ,
    \nonumber
 \\
\mbox{thus for (1)}  
 &    &  
     \hspace{5mm}
 \delta K  \; = \; 375 \: \mbox{kg (m/s)${}^2$} \:  .
    \nonumber
\end{eqnarray}

For (2) we need some preliminary calculations: the temperature at the upper limit of $M_1$ in the initial state is $T_{h1} = 279.53$~K, the pressure at the interface and the stamp is
\begin{eqnarray}
 p_{i}  &  = &  
    760 \;  
           \left( \;
           \frac{299.4}{307.8}
       \; \right)^{g/(R \: b)} 
 \;  =  \; \;
   606.828 \: \mbox{mm-Hg}
   \:  ,
    \nonumber  \\
 p_{h}  &  = &  
    p_i \;  
           \left( \;
           \frac{279.53}{299.4}
       \; \right)^{g/(R \: b)} 
 \;  =  \; \;
   479.192 \: \mbox{mm-Hg}
   \:  ,
    \hspace{5mm}
    p_0 \: - \: p_i \: + \:  p_h \; = \; 632.364 \: \mbox{mm-Hg} \: .
    \nonumber 
\end{eqnarray}

Now, the new temperature values at the boundaries of $M_1$ comes from the equation of the adiabatic state:
\begin{equation}
    299.4 \;  
           \left( \;
           \frac{760}{606.828}
       \; \right)^{\varkappa} 
 \;  =  \; \;
   319.65 \: \mbox{K} \: ,
     \hspace{10mm}
    279.53 \;  
           \left( \;
           \frac{632.364}{479.192}
       \; \right)^{\varkappa} 
 \;  =  \; \;
   303.01 \: \mbox{K} \: .
    \nonumber 
\end{equation}

The quotient of the difference of these temperatures by $A$ gives the height of the mass $M_1$, when nearly the whole $M_2$ has overlaid.

We have to compute $\delta K$ from ($5_1$) with $m_2$, $a$, $b$ as above and
\begin{equation}
    h 
 \;  =  \; \;
   1675 \: \mbox{m} \: ,
     \hspace{10mm}
    T_2 
 \;  =  \; \;
   307.8 \: \mbox{K}    \: ,
     \hspace{10mm}
    T_{01}
 \;  =  \; \;
   319,65 \: \mbox{K}    \: ,
    \nonumber 
\end{equation}
and for (2):
\begin{equation}
 \delta K  \; = \; - \:  371 \: \mbox{kg (m/s)${}^2$} \:  ,   \nonumber 
\end{equation}
which is a value of nearly the same value as for (1), but of opposite sign.
Thus, the layer $j$ will be searched for in the middle.

To (3). Assuming $\alpha \: M_2$ is the mass, which was initially situated between the altitudes $1000$ and $2000$~m, we calculate $\delta K$  for $1$~kg from the layer at the height $1000$~m.
After the necessary preliminary work, the following is to be used in ($5_1$):
\begin{equation}
    h 
 \;  =  \; \;
   1827 \: \mbox{m} \: ,
     \hspace{10mm}
    T_2 
 \;  =  \; \;
   303.6 \: \mbox{K}    \: ,
     \hspace{10mm}
    T_{01}
 \;  =  \; \;
   309.4 \: \mbox{K}    \: ,
    \nonumber 
\end{equation}
so that for (3):
\begin{equation}
 \delta K  \; = \; - \:  40 \: \mbox{kg (m/s)${}^2$} \:  ,   \nonumber 
\end{equation}
which is also a negative value.
The area $j$ is at the height of about $1100$~m.
Only from the top 900 m of $M_2$ mass can ascend with increasing kinetic energy of the system

The potential energy of our system is calculated by using the following method: 
$1$~kg of the uppermost layer of $M_2$ yields $\delta K = 380$, 
$1$~kg from the height $1000$~m  yields $\delta K = 0$ and 
every kilogram from the range between $1000$ and $2000$~m creates an amount $\delta K = 190$~kg~(m/s)${}^2$.
The mass of this altitude interval compares with $M_1 + M_2$ like $p_{1000}-p_{2000}$ compare with $p_0-p_h$, or as $72.8$ compares with $280.8$.
If the {\it available kinetic energy\/} is evenly distributed into $M_1 + M_2$, the maximum amount of $190 \times 72.8 / 280.8$, i.e. close to $50$~kg~(m/s)${}^2$, corresponding to the speed $V = 10$~m/s.



\vspace{3mm}
$\bullet$ {\bf 11.} Now we can look at a model which is quite often considered.

We choose the representation of Helmholtz \footnote{Lectures and speeches, fifth edition, II, 155. (Braunschweig 1903.)}. A whirlwind in the tropics is to arise in the following way:

Initial state: 
a vast mass of steam-saturated air in stable equilibrium and with absolute calmness, above a dry air and at the limit a somewhat smaller density.

Beginning of the movement and increase of the storm: 
a small impulse, or an accidental pressure decrease, suffices to let the humid air to rise at a point with free buoyancy; in that place a column is filled with lighter air and the pressure sinking on the ground; moist mass flows in spirals; the effect of the centrifugal force increases the horizontal pressure gradient in the lower layers, and the speed of the inflowing air increases as the compressive forces work.
The movement is maintained as long as moist air can rise in the tubular region.
The inflow comes from a region which is larger than the actual storm area and the ascending moist mass spreads over a large surface.

Extinguishing of the storm:
the steam-saturated layer on the ground gradually becomes shallower, the dry mass sinks, the decrease in heat is about of $1$~C for a $100$~m, while the temperature gradient in the moist mass is about $0.4$~C/100m.
This reduces the difference in density between the air in the tube and outside it at the same level.
The driving force of the storm is extinguished by the buoyancy and the horizontal pressure gradients vanish.
The {\it living force\/} is consumed by friction and mixture.

Final state: a part of the moist mass still lies below the dry air above; the other part of the humid air forms the third layer; the higher layers, which had no part in the whole process, are as in the initial state.

With reference to this model, the final example is computed. 
From the initial state, a greatest possible increase in kinetic energy can be obtained.
If we think of an extended air mass of the same temperature and vapour concentration instead of the unitary column assumed there, we have an initial state similar to the one posited in the Ferrel-Helmholtz model.

The storm energy is favoured  because of the unstable condition at the boundary surface, also because of the greatest vertical temperature gradient in the dry mass, and mainly because of the high temperature and the large steam content in the lower part.
At the same time, a larger $V$ could be obtained if the height of the upper mass had been set higher than $2000$~m.

The {\it available kinetic energy\/} of the system is such that each kilogram of the reacting masses could reach the velocity of $10$~m/s and the moderate speed of $30$~m/s in tropical storms could only arise in the ninth parts of the whole mass.

The energy supply to the system is smaller than that of the two chamber system with $10$~C of temperature and $2000$~m altitude differences.

\begin{figure}[hbt]
\centering
\includegraphics[width=0.9\linewidth,angle=0,clip=true]{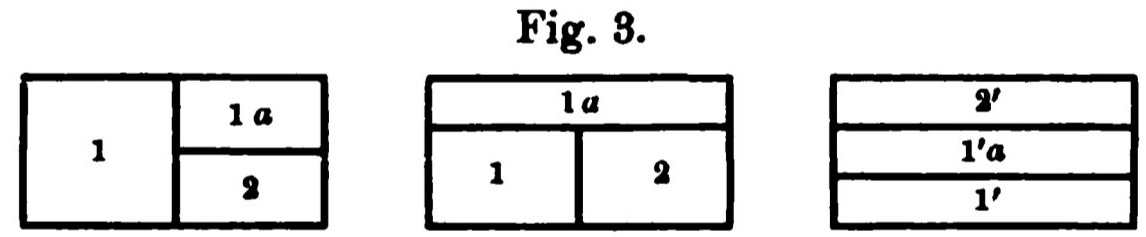} \\
\hspace{1cm} (A) \hspace{4cm}  (B) \hspace{4cm} Final State
\label{fig_3}
\end{figure}

\vspace{3mm}
$\bullet$ {\bf 12.} Combine horizontal temperature drop and a rising steam-rich stream.

We must not confine ourselves to the consideration of two masses.
We want to include three of them in our system.

(Fig.3): 1 is the potentially coldest dry air (on the left); 1a is initially above the vapour-rich mass 2 in vertically stable stratification, so that even large impulses would not be sufficient to drive through the initial position 2 through 1a.
But as soon as the mass 1 spreads below 2, the vapour-rich air is raised, expanded and condensed, it can break through the overlying dry mass, and the final layer will be 1', 1'a, 2' (on the right).
A part of the potential energy of this system lies in the temperature differences between 1 on the one hand, 2 on the other and 1a on the other hand on the same level in case (A), between 1 and 2 in case (B). 
The remaining part in the vapour content of 2.

We have so far been dealing with quite abstract considerations.
If we wish to make use of the results, we must first know whether such a condition, as assumed in Fig.~3, precedes the storm, or whether the initial state postulated by the Helmholtz model is the same for a storm region before the storm (a steam-saturated layer of at least $1000$~m height on a ten-fold larger surface, above dry air a few kilometres high with the gradient of about $0.7$~C/100m).

Perhaps it will be better to judge this after a few years, when the balloon and kite climbs are continued over tropical seas that have only just begun so happily.
For the time being, I would assume that the potential energy, which is dependent on the horizontal temperature distribution, and which appears in our latitudes as the chief source of the storm energy, does not disappear even in the tropics, as opposed to the portion of condensation heat convertible into kinetic energy.

There is no lack of evidence.
In northern India, even up to $20$ degrees of latitude, stormy areas of much greater size are known than the Bengali cyclones. In terms of barometer depression, speed and direction of progress, in the wind force, they are similar to the storm regions of Northern and Central Europe \footnote{Cold Weather Storms in India, 1876-1891. Indian Meteor. Memoirs 4, 529.}
Indications of squills in Bengali cyclones are also found in various parts of the Indian Memoirs, as well as those occurring near the equator.
On account of lack of time, I recall, according to reports from the shipping captain, some still older data from Reye (Hurricanes p.159).

- Madras Masalipatam - Cyclones, 21st to 23rd May 1843: the strongest winds were at $1$~UTC, then there were hard and intermittent shocks accompanied by great and terrible heat. After the storm had turned to SW, there were alternate hot and cold wind blows.

- October Cyclones 1848: hot and cold gusts of wind are felt distinctly.

- Madras-Sturm 1836 (11 to 12 NB.): an extraordinary change occurred suddenly in the temper of the air.

The last report reads like a report from our regions.
But the rapid alternation of cold and warm winds is not unknown to us \footnote{Compare, for example, the thermograms of Vienna, January 23, 1892; Vienna, 9th to 10th of February, 1896, in the Met., 1898, p. 12; also those of Pressburg, January 16, 1899; St. Polten, 18-19 December 1899; in the yearbooks of the K. K. Central Institution for Met, 1899, v., P. 10 and 15.}. 
One finds it close to the jump of temperatures in the boundary layer.
According to this, the main source of the kinetic energy could be the same for tropical cyclones as for our storms.
The transformation of a small part of the heat of the steam into a vigorous force would come to a greater degree than in the middle latitudes, but only by the way.
I do not know the circumstances in the tropics from my own experience and I would like to mention this opinion with caution.

On the same page of Reyes's book, taken from the last quotations, one finds information about ice-cold rains in grains near the equator.
The drops must have been placed at a very high altitude, so as to fall off so coldly.
Reye says the cold wind blows are air cooled by rain or hail. 
This too is to be considered as to whether precipitation, which occurs during the storm, can lead to the storm by cooling a part of the air mass.
[I know of a very severe hail-fall (Vienna, June 7, 1894, Met. Ztschi, 1898, p. 15) with a very slight lowering of the temperature of the lower air layer]
Regardless of any assumption as to the causes of the cold and hot gusts of wind, those messages teach that the information about uniform or symmetrical temperature distribution in tropical cyclones is to be suspected.

\vspace{5 mm}
\noindent {\Large \bf \underline{Supplement: On a possible development of storm theory}.}
\vspace{2 mm}

An idea of the energy source of the storm regions with asymmetrical temperature distortion can be obtained from this roughest model.
But for the question of how the storm arises, or in the picture, how the door opens, the chambers with a dormant air mass will not be needed.

Storm is comparatively rare for horizontal temperature drops and for masses with a large supply of potential energy everywhere.
The masses remain at a moderate rate in an almost stationary state, which is obtained either by heat processes, with vertical circulation components, or even dynamically.
The conditions of the stability of this state are known.
Disturbances of a certain small quantity do not lead to a stormy movement. They can propagate in waves, without swelling, and they would soon be extinguished by friction.

Are there any major disturbances of the dynamic equilibrium that grow by themselves by drawing from the potential energy, perhaps from the supply of kinetic energy?
This seems to me the analytical problem of the formation of a storm. 
The balance eats the wall and the disturbance the door that opens between the chambers (Dambreak).

For such an investigation, the search for productive energy sources was a not useless preliminary work.
Sometimes the conjecture or assertion is made that the storm arises between regions of opposing winds.
Tropical cyclones are also to form between the trades and monsoon regions (see Thom, Reye, I, p. 148).

If we wish to trace this analytically, we can only observe the direction and velocity of the air currents, then the supplies from which the disturbance might fatten, the kinetic energy of the undisturbed system, and the very small potential energy of the pressure distribution.
Or one introduces considerable horizontal temperature differences, and thus another very productive source.

In order to make the calculation as short as possible, one attempts to unite the essential conditions of the task in the simplest model.
I would first choose anti-parallel straight-line currents of incompressible fluids as an undisturbed state in the rotating system with flat levelling surfaces (Met. Ztsohr. Hann-Band, S. 244), or a current next to a quiescent mass.
What disturbance is properly introduced cannot be guessed from the outset.
This can be a patience test according to the axiom of Ostwald: ``the simplest one always comes last''.

Meteorologists will appear to be abstract.
It would certainly be desirable to have some guidance from experience.
But an account cannot be derived from the usual Isobar cards.
Such questions, how, in our gusts, masses of very different temperatures come directly together? how the steep boundary surface arises? must be answered empirically.
These include two-consciousness-conscious observations over a large area, even for phenomena of relatively small extent. 
For large storm regions, for example depression in the Adriatic, sufficient observations are even more difficult to obtain.
It was recently thought that the pressure of cold air masses to Italy was caused by pressure distribution.
We now believe that a depression disappears if it is not nourished by the difference in temperature between the reacting air masses.
The interaction between pressure distribution and flow is, as in the aerodynamic calculations, also as the representation of observed phenomena, the source of many difficulties.
I find it scarcely possible to use the observations alone as a useful model of the development of the storm.
{\it A computer with sufficient knowledge of the observations, with imagination and a lot of patience, may reach the goal\/}.  \footnote{underlined by the translator}.

Vienna, June 1906.


\vspace{5mm}
\noindent{\Large\bf \underline{References}}
\vspace{2mm}

\noindent{$\bullet$ Margules, Max} {(1903-1905)}.
\"Uber die Energie der St\"urme (On the energy of the storms)
{\it Jahrb\"ucher der K K. Zentralanstalt f\"ur Meteorologie und Erdmagnetismus, Jahrgang 1903, Wien 1905
(Yearbook 1903 of the K K. Central Institute for Meteorology and Earth magnetism, Vienna, 1905)\/},
{\bf Vol. 40}: p.1--26.
See also the Translation by C. Abbe in the {\it Smithsonian Miscellaneous collections,} {\bf Vol. 51} (Issue 4): p.533--595, 1910.

\noindent{$\bullet$ Margules, Max} {(1906)}.
Zur Sturmtheorie.
{\it Meteorologische Zeitschrift\/}
(Meteorological journal; Journal of 
the K.K. Austrian Society of Meteorology)
{\bf Vol. 23} (Issue 11): p.481--497. 
\url{https://archive.org/details/meteorologische01unkngoog}

  \end{document}